\documentstyle[prl,twocolumn,aps]{revtex}
\input{epsf.tex}
\newcommand{\bm}[1]{\mbox{\boldmath$#1$}}
\newcommand{\up}{\uparrow}
\newcommand{\dn}{\downarrow}
\newcommand{\tfrac}[2]{{\textstyle\frac{#1}{#2}}}
\newcommand{\Vpara}{V_{\parallel}}
\newcommand{\Vbot}{V_{\bot}}
\begin{document} \draft

\twocolumn [ \hsize\textwidth\columnwidth\hsize\csname
@twocolumnfalse\endcsname
\title{Exact Bond-Located Spin Ground State in the Hubbard
Chain with Off-Diagonal Coulomb Interactions}
\author{Kazuhito Itoh$^1$, Masaaki Nakamura$^2$, and Norihiro
  Muramoto$^1$}
\address{$^1$Institute for Solid State Physics, University of Tokyo,
  Kashiwanoha, Kashiwa-shi, Chiba 277-8581, Japan}
\address{$^2$Max-Planck-Institut f\"{u}r Physik komplexer Systeme,
  N\"{o}thnitzer Stra{\ss}e 38, 01187 Dresden, Germany}
\date{\today}
\maketitle
\begin{abstract}
  \widetext\leftskip=0.10753\textwidth \rightskip\leftskip
  We show the existence of an exact ground state in certain parameter
  regimes of one-dimensional half-filled extended Hubbard model with
  site-off-diagonal interactions.  In this ground state, the
  bond-located spin correlation exhibits a long-range order.  In the
  case when the spin space is SU(2) symmetric, this ground
  state degenerates with higher spin states including a fully
  ferromagnetic state.  We also discuss the relation between the exact
  bond-ordered ground state and the critical bond-spin-density-wave phase.
\end{abstract}
\indent
]\narrowtext

The Hubbard model is one of the generic models to describe interacting
electrons in narrow-band systems~\cite{Hubbard}.  The on-site repulsion
of this model is due to the matrix elements of the Coulomb interaction
corresponding to the on-site Wannier states, and the other matrix
elements are neglected.  The importance of the neglected
nearest-neighbor exchange interactions was, however, stressed and
discussed for stabilization of ferromagnetism or a dimerized
state~\cite{Hirsch,CGL}.

Japaridze first discussed using the weak-coupling theory the possibility
of the bond-spin-density-wave (BSDW) ground state connected with the
site-off-diagonal nature of the pair-hopping term~\cite{Japaridze}.  He
introduced the BSDW order parameters ${\bm {\cal O}_{i}} = ({\cal
O}_{i}^{\mathstrut x}, {\cal O}_{i}^y, {\cal O}_{i}^{\mathstrut z})$:
\begin{equation}
  {\bm {\cal O}}_{i} = \tfrac{1}{2}(-1)^{i}\sum_{\sigma\sigma'}
  (c_{i\sigma}^{\dagger} 
  \bm{\tau}_{\sigma\sigma'}^{\mathstrut} 
  c_{i+1\sigma'}^{\mathstrut} + {\rm h.c.})
\end{equation} 
with the Pauli matrices $\bm{\tau} = (\tau^{x}, \tau^{y}, \tau^{z})$,
where the operator $c_{i\sigma}^{\dagger}$ ($c_{i\sigma}^{\mathstrut}$)
creates (annihilates) an electron with spin $\sigma$ ($=\up,\dn)$ at the
$i$-th site.  The $z$-component of these order parameters ({BSDW-$z$})
${\cal O}_{i}^{z}$ describes a staggered magnetization with spins
located on bonds between adjacent sites (see Fig.~\ref{fig:BNeel}).

\begin{figure}
 \begin{center}
  \epsfxsize=2.5in \leavevmode \epsfbox{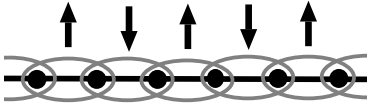}
 \end{center}
\caption{The BN\'eel state given by Eq.~(\protect{\ref{eqn:twf}}). 
The enclosed pairs of sites (gray ovals) indicate electron-hole dimers.}
 \label{fig:BNeel}
\end{figure}

First in this Letter, we exactly demonstrate that the half-filled
one-dimensional (1D) Hubbard model with site-off-diagonal interactions
and a spin anisotropy has a doubly degenerate bond-ordered ground state
in an intermediate coupling regime, using the optimal ground state
approach~\cite{Majumder-G,Boer-S,KSZ1,Kolezhuk-M,KI}.  This bond-ordered
state is interpreted as a N\'eel state of bond-located spins, and the
state realizes the fully BSDW long-range order with respect to the
z-direction.  For this reason we call the state the {\it Bond-N\'eel }
(BN\'eel) state in the following.  In this state there are no
correlations with respect to both of site-located charge and spin
density operators except for the nearest neighbor.  The system
possesses, therefore, charge and spin gaps between the ground state and
excited states.

Next we consider a spin SU(2) symmetric case, and we show that the
existence of the phase in which the BSDW correlation is dominant by
using numerical methods as the level-crossing
approach~\cite{Nakamura,Nakamura-I-M}.  The BSDW correlation shows a
power-law decay at large distances and the system has a gapless spin
excitation.  This BSDW phase borders on the BN\'eel phase on the
multicritical point.  We discuss the relation between the BN\'eel state
and the critical BSDW state.

We consider the 1D extended Hubbard model with on-site and
nearest-neighbor interaction given by
\begin{displaymath}
  \tfrac{1}{2}\sum_{ijkl}\sum_{\sigma\sigma'}
  \langle{ij}|v_{\sigma\sigma'}|{kl}\rangle
  c_{i\sigma}^{\dagger} c_{j\sigma'}^{\dagger}
  c_{l\sigma'}^{\mathstrut} c_{k\sigma}^{\mathstrut},
\end{displaymath}
where the matrix elements with respect to on-site or nearest-neighbor
only are nonzero.  The Hamiltonian is ${\cal
H}=\sum_{\langle{i,j}\rangle}h_{ij}$, where the local Hamiltonian
$h_{ij}$ is given by
\begin{eqnarray}
  h_{ij} &=& 
  \sum_{\sigma}
  (c_{i\sigma}^{\dagger} c_{j\sigma}^{\mathstrut} + {\rm h.c.})
  \bigl[-t + X(n_{i-\sigma} + n_{j-\sigma})\bigr]
  \nonumber\\
  && \vphantom{\sum_\sigma}
  + \tfrac{1}{2}U
  \bigl[(n_{i\up} - \tfrac{1}{2})(n_{i\dn} - \tfrac{1}{2})
  + (n_{j \up} - \tfrac{1}{2})(n_{j\dn} - \tfrac{1}{2})\bigr]
  \nonumber\\
  && {} + \sum_{\sigma\sigma'} V_{\sigma\sigma'} 
  (n_{i \sigma}-\tfrac{1}{2})
  (n_{j \sigma'}-\tfrac{1}{2})
  \nonumber\\
  && {} +
  \tfrac{1}{2}W\sum_{\sigma\sigma'}
  (c_{i\sigma}^{\dagger} c_{j\sigma}^{\mathstrut}
  + c_{j\sigma}^{\dagger} c_{i\sigma}^{\mathstrut})
  (c_{i\sigma'}^{\dagger} c_{j\sigma'}^{\mathstrut}
  + c_{j\sigma'}^{\dagger} c_{i\sigma'}^{\mathstrut}).
  \label{eqn:ghub-l}
\end{eqnarray}
Here the summation is taken over pairs of neighboring sites
$\langle{i,j}\rangle$, and the number operators is defined as
$n_{i\sigma} = c_{i\sigma}^{\dagger} c_{i\sigma}^{\mathstrut}$.  The
first term in the Hamiltonian (\ref{eqn:ghub-l}) is the single-particle
hopping ($t$-term) including the bond-charge interaction $X =
\langle{ii}|v_{\sigma\sigma'}|{ij}\rangle$.  The on-site coupling is $U
= \langle{ii}|v_{\sigma,-\sigma}|{ii}\rangle$.  Nearest-neighbor density
interaction with opposite and parallel spins are parametrized by
$V_{\sigma,-\sigma} = \Vbot=\langle{ij}|v_{\sigma,-\sigma}|{ij}\rangle$
and $V_{\sigma\sigma} = \Vpara =
\langle{ij}|v_{\sigma\sigma}|{ij}\rangle$, respectively.  Here we
introduce spin anisotropy due to some factors, such as the crystal
structure and spin-orbit interaction. The bond-bond interaction is
controlled by $W = \langle{ij}|v_{\sigma\sigma'}|{ji}\rangle =
\langle{ii}|v_{\sigma\sigma'}|{jj}\rangle$.

Now let us show that the above Hamiltonian can be factorized in some
parameter space.  For this purpose, we introduce a local bond operator
with spin $\sigma$ defined by
\begin{equation}
  R_{ij}^{\sigma}(\lambda,\mu)
  = \lambda\, a_{ij\sigma}^{\dagger} a_{ij\sigma}^{\mathstrut}
  + \mu\, s_{ij\sigma}^{\mathstrut} s_{ij\sigma}^{\dagger},
  \label{eqn:Raass}
\end{equation}
where the operators $s_{ij\sigma}^{\dagger}$ and
$a_{ij\sigma}^{\dagger}$ denote the symmetric and antisymmetric
bond-creation operators, respectively: $s_{ij\sigma}^{\dagger} = \bigl(
c_{i\sigma}^{\dagger} + c_{j\sigma}^{\dagger} \bigr)/\sqrt{2}$,
$a_{ij\sigma}^{\dagger} = \bigl( c_{i\sigma}^{\dagger} -
c_{j\sigma}^{\dagger} \bigr)/\sqrt{2}$.  The bond operator
$R_{ij}^{\sigma}(\lambda,\mu)^{\mathstrut}$ can be rewritten by using
the operators $B_{ij}^{\sigma} =
c_{i\sigma}^{\dagger}c_{j\sigma}^{\mathstrut} + {\rm h.c.}$ and
$N_{ij}^{\sigma} = n_{i\sigma}+n_{j\sigma }$:
\begin{equation}
  R_{ij}^{\sigma} (\lambda,\mu)
  = \frac{\lambda+\mu}{2} \bigl( 1-B_{ij}^{\sigma} \bigr)
  + \frac{\lambda-\mu}{2} \bigl( N_{ij}^{\sigma}-1 \bigr).
\end{equation}
The local Hamiltonian (\ref{eqn:ghub-l}) can be written as follows:
\begin{eqnarray}
  h_{ij} &=&
  \vphantom{\frac{U}{2}}
  R_{ij}^{\up}(\alpha,\beta)\, R_{ij}^{\dn}(\alpha',\beta')
  +R_{ij}^{\up}(\lambda,\mu)\, R_{ij}^{\dn}(\lambda',\mu')
  \nonumber\\ 
  && {} + \sum_{\sigma\sigma'} \tilde{V}_{\sigma\sigma'}
  (n_{i \sigma}-\tfrac{1}{2})(n_{j \sigma'}-\tfrac{1}{2}) 
  \label{eqn:local}\\ 
  && {} + (W+X-t)\sum_{\sigma }B_{ij}^{\sigma}
  + X(n_{i} + n_{j}) -(t+X),\nonumber 
\end{eqnarray}
where $\tilde{V}_{\sigma\sigma} = \Vpara-W$ and
$\tilde{V}_{\sigma\sigma'} = \Vbot-U/2$ ($\sigma \ne \sigma'$).  Here
the parameters of operators $R_{ij}^{\sigma}$ in (\ref{eqn:local}) are
constrained to satisfy the relations:
\begin{eqnarray}
  2(\alpha\alpha' + \lambda\lambda')
  &=& U + 6W - 4t, \nonumber\\
  2(\beta\beta' + \mu\mu')
  &=& U - 2W + 4t, \\ 
  2(\alpha\beta' + \lambda\mu')
  &=& 2(\alpha'\beta + \lambda'\mu)
  = -U + 2W. \nonumber 
\end{eqnarray} 
When $\lambda,\mu \ge 0$, we see from the Eq.~(\ref{eqn:Raass}) that the
bond operator is positive semidefinite: $\bigl\langle
R_{ij}^{\sigma}(\lambda,\mu) \bigr\rangle \ge 0$.  Then for $U =
2\Vbot$, $W = \Vpara = t-X$ the local operator ${\tilde h}_{ij} =
h_{ij}- X(n_{i} + n_{j}) + t+X$ is also positive semidefinite as long as
$\alpha, \alpha', \beta, \beta', \lambda, \lambda', \mu, \mu' \ge 0$.

From these facts, we find that the energy of the global Hamiltonian
${\cal H}$ is bounded from below in the following regime:
\begin{eqnarray}
  && U = 2\Vbot, \quad
  W = \Vpara = t-X, \nonumber\\
  && \max\{2W-4t,\ -6W+4t\} \le U \le 2W.
\label{eqn:regime}
\end{eqnarray}
We denote the global Hamiltonian ${\cal H}$ in this regime as ${\cal
H}_{\rm b}$.  We impose the periodic boundary condition on the system
and choose the number of lattice sites $L$ to be even in the following.
The lower bound is given as $E_{\rm lower}=-(t+X)L+2XN$, where $N$
denotes the total number of electrons.  To obtain an upper bound of the
ground-state energy, we consider the following trial wave functions:
\begin{equation}
  \bigl|\Psi_{0}^{\sigma}\bigr\rangle = \hspace{-.6em}
  \prod_{i\in\{{\rm odd}\}}\hspace{-.6em} s_{i,i+1,\sigma}^{\dagger}
  \hspace{-.6em}
  \prod_{j\in\{{\rm even}\}}\hspace{-.8em} s_{j,j+1,-\sigma}^{\dagger} 
  |0\rangle, 
  \label{eqn:twf}
\end{equation}
where $|0\rangle$ denotes the vacuum.  A schematic illustration of this
state is given in Fig.~\ref{fig:BNeel}.  The state corresponds to a
half-filled state with a density $N/L = 1$.  Since
$R_{ij}^{\sigma}(\lambda,\mu)\, s_{ij\sigma}^{\dagger}|0\rangle = 0$,
the state $s_{ij\sigma}^{\dagger}|0\rangle$ is an eigenstate of the
local Hamiltonian ${\tilde h}_{ij}$ with eigenvalue zero: ${\tilde
h}_{ij}^{\mathstrut} s_{ij\sigma}^{\dagger}|0\rangle = 0$.  Rewriting
the global Hamiltonian as ${\cal H}_{\rm b} = \sum_{\langle i,j \rangle}
\bigl[\, {\tilde h}_{ij}+X(n_i+n_j)-(t+X) \bigr]$, we can see that the
state (\ref{eqn:twf}) is also an eigenstate:
\begin{equation}
  {\cal H}_{\rm b}\, \bigl|\Psi_{0}^{\sigma}\bigr\rangle
  = -(t-X)L\, \bigl|\Psi_{0}^{\sigma}\bigr\rangle.
\end{equation}
This means that the ground-state energy is bounded from above at $E_{\rm
upper} = -(t-X)L$.  At half filling, therefore, the upper and lower
bound coincide, and the eigenstate $\bigl|\Psi_{0}^{\sigma}\bigr\rangle$
is the ground state in the regime~(\ref{eqn:regime}).  The ground-state
energy is given as $E = -(t-X)L$.

\indent

\begin{figure}
\begin{center}
\noindent \epsfxsize=2.75in \leavevmode \epsfbox{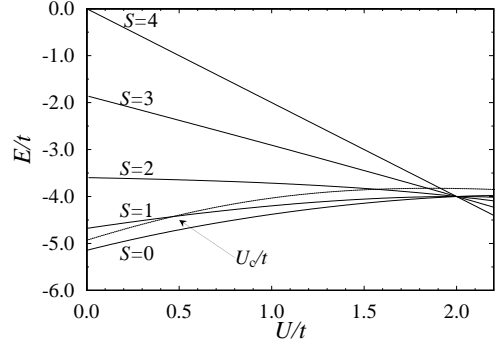}
\end{center} 
\caption{Energies in the $L=N=8$ system at $V=W=t$. In the region where
the singlet excitation (dotted line) is lower than the triplet one, the
spin excitation has a gap.  The singlet-triplet level crossing point
corresponds to the CDW-BSDW phase boundary. }
\label{fig:spin_exs}
\end{figure}

When $U = 2W \ge t$, the system is in the the regime~(\ref{eqn:regime}),
and the spin exchange interaction obtains the SU(2) symmetry, $\Vpara =
\Vbot =V$.  Then the operators $S_{\rm tot}^\alpha = \sum_{i}
S^\alpha_{i}$, $\alpha = +,-,z$ commute with the Hamiltonian.  In this
case, we can deduce that the ground state
$\bigl|\Psi_{0}^{\sigma}\bigr\rangle$ and a fully polarized
ferromagnetic (FM) state, $|{\rm FM}\rangle =
\prod_{i}\,c_{i\up}^{\dagger}\,|0\rangle$, are degenerate.  Since
$S_{\rm tot}^{+}\bigl|\Psi_{0}^{\sigma}\bigr\rangle$ does not vanish,
the states
\begin{equation}
  \biggl|
  {l_1\atop m_1}{l_2\atop m_2}{\dots\atop\dots}
  \biggr\rangle^{\sigma}
  = \prod_{k}\, \bigl(S_{\rm tot}^{+}\bigr)^{{l}_{k}}
  \bigl(S_{\rm tot}^{-}\bigr)^{{m}_{k}}\, 
  \bigl|\Psi_{0}^{\sigma}\bigr\rangle
  \label{eqn:d-states}
\end{equation}
are also eigenstates with the lowest energy eigenvalue and with $S_{\rm
tot}^{z} = \sum_{k}(l_k - m_k)_{\mathstrut}$, ($|S_{\rm tot}^{z}| \le
L/2$) as long as $\bigl|{l_1\atop m_1}{l_2\atop m_2}{\dots\atop\dots}\,
\bigr\rangle^{\mathstrut\sigma}_{\displaystyle\mathstrut} \neq 0$.
These states, however, may not be linearly independent each other. The
eigenstates of $\bm{S}_{\rm tot}^{2}$ with $S_{\rm tot}^{z}$ may be
constructed by linear combinations of the states~(\ref{eqn:d-states})
with $S_{\rm tot}^{z}$. Then these $\bm{S}_{\rm tot}$ states $(|\bm
S_{\rm tot}| \le L/2)$ may degenerate.  In Fig.~\ref{fig:spin_exs} we
show energies in the $L=8$ system at $W=V=t$ in Eq.~(\ref{eqn:ghub-l})
using the exact diagonalization.  At the exactly solvable point
($U=2t$), all the lowest states with total spin $|\bm S_{\rm tot}|\leq
L/2$ and the density $N/L=1$ degenerate.

\begin{figure}
 \begin{center}
  \epsfxsize=2.7in \leavevmode \epsfbox{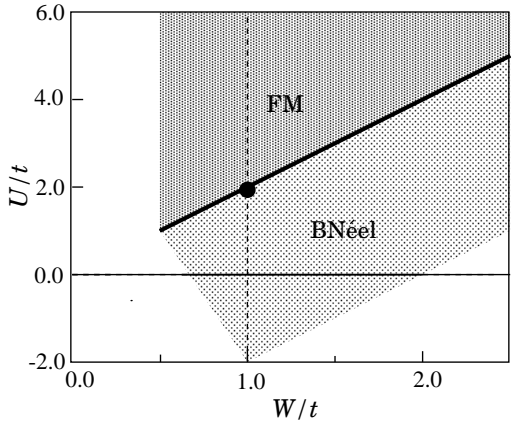}
 \end{center}
\caption{The phase diagram of the present model
 (\protect{\ref{eqn:ghub-l}}) in the $U$--$W$ plain.  The parameters of
 the model are fixed as $U=2\Vbot$ and $W=\Vpara=t-X$. The $X$-term
 vanishes on the vertical dotted line. The point denotes at $\Vbot =
 \Vpara, X=0$.}
 \label{fig:pd1}
\end{figure}

For $\Vbot > \Vpara$, i.e., $t \le 2W < U$, the Hamiltonian can be
expressed as
\begin{equation}
{\cal H}
  = {\cal H}_{\rm b}\bigr|_{U=2W}
  + \gamma \sum_{\langle i,j \rangle}
  \bigl(N_{ij}^{\up} - 1\bigr) \bigl(N_{ij}^{\dn}- 1\bigr)
\end{equation}
with $\gamma = U/2-W > 0$.  The ground state is the FM state, since
$|{\rm FM}\rangle$ gives the lowest eigenvalue both of ${\cal H}_{\rm
b}\bigr|_{U=2W}$ and $\sum_{\langle{i,j}\rangle}
\bigl(N_{ij}^{\up}-1\bigr) \bigl(N_{ij}^{\dn} - 1\bigr)$.  The system
undergoes, therefore, a first-order phase transition at this
level-crossing point $\gamma = 0$. At $U=-2W=-2t$, $X=0$, the
$\eta$-paring operator with momentum $\pi$, $\eta^{\dagger} =
\sum_{j}(-1)^{j}c_{j\downarrow}^{\dagger}c_{j\uparrow}^{\dagger}$,
commutes with the Hamiltonian. Therefore the $\eta$-paring state:
$|\eta\rangle = (\eta^{\dagger})^{N/2} |0\rangle$ is also an exact
ground state, which agrees with the result by de Boer and
Schadschneider~\cite{Boer-S}.  From the above discussion, we obtain the
phase diagram of the model (\ref{eqn:ghub-l}) as shown in
Fig.~\ref{fig:pd1}.

\indent

In order to analyze the properties of two ground states
$\bigl|\Psi_{0}^{\up}\bigr\rangle$ and
$\bigl|\Psi_{0}^{\dn}\bigr\rangle$ in the regime~(\ref{eqn:regime}), we
introduce the matrix product representation~\cite{KSZ1,Kolezhuk-M}.
This is useful for calculating the ground state expectation values and
correlation functions.  First, we describe these states in terms of
products of local $2 \times 2$ matrices as,
\begin{equation}
  \bigl|\Psi_{0}^{\sigma}\bigr\rangle =
  2^{-L/2}\: {\rm Tr}\: g_{1}^{\sigma} \otimes g_{2}^{-\sigma} \otimes
  \cdots \otimes g_{L-1}^{\sigma} \otimes g_{L}^{-\sigma},
\end{equation}
where two matrices $g_{i}^{\sigma}$, $\sigma = \up,\dn$, are given by
\begin{equation}
  g_{i}^{\sigma} =
  \left(
    \begin{array}{cc}
      |{\sigma}\rangle_{i} & |2\rangle_{i}\\
      |0\rangle_{i} & |{-\sigma}\rangle_{i}
    \end{array}
  \right).
\end{equation}
Here $|\sigma\rangle_{i} = c_{i\sigma}^{\dagger}|0\rangle_{i}$,
$|2\rangle_{i} = c_{i\up}^{\dagger}c_{i\dn}^{\dagger}|0\rangle_{i}$, and
$\otimes$ denotes the usual matrix multiplication of $2 \times 2$
matrices with a tensor product of the matrix elements.  Note that
${\tilde h}_{ij} \bigl(\, g_{i}^{\sigma} \otimes g_{j}^{-\sigma} \bigr)
= 0$.  Next, we introduce $4 \times 4$ transfer matrices:
$G_{l_{1}l_{2}}^{\sigma\sigma'} \equiv
g_{n_{1}n_{2}}^{\sigma\dagger}\,g_{m_{1}m_{2}}^{\sigma'}$, where
$l_k=(n_k,m_k)$ and the indices correspond as $l_k=1,2,3,4
\leftrightarrow (n_k,m_k)=(11),(12),(21),(22)$, respectively.  Since the
overlap between $\bigl|\Psi_{0}^{\dn}\bigr\rangle$ and
$\bigl|\Psi_{0}^{\up}\bigr\rangle$ for size $L$ is evaluated by
$\bigl\langle\Psi_{0}^{\up}\big|\Psi_{0}^{\dn}\bigr\rangle = 2^{-L}\:
{\rm Tr}\,(G^{\up \dn})^{L} = 4 \!\cdot\! 2^{-L}$, these two states are
orthogonal in the limit $L \to \infty$.  The ground state is, therefore,
doubly degenerate, and the translational symmetry of the system is
spontaneously broken except for $U=2W$.

The two-site correlation functions of operators $A_{1}$ at the 1st site
and $B_{r}$ at the $r$-th site can be calculated by $\langle A_{1}B_{r}
\rangle = {\rm Tr}\,Z(A)\,G^{r-2} Z(B)\,G^{L-r}$ with $Z(A) =
g^{\sigma\dagger} A\, g^{\sigma}$. Thus, the nearest neighbor
correlations are obtained as follows, $\langle n_{i} n_{i+1} \rangle
=2$, $\bigl\langle S_{i}^{z} S_{i+1}^{z} \bigr\rangle = -\frac{1}{4}$,
and $\bigl\langle S_{i}^{+} S_{i+1}^{-} \bigr\rangle = 0$.  On the other
hand, the two-point charge and spin correlation functions for $|i-j| \ge
2$ are $\langle n_{i} n_{j} \rangle - \langle n_{i} \rangle \langle
n_{j} \rangle = 0$ and $\bigl\langle S_{i}^{z} S_{j}^{z} \bigr\rangle
=0$, respectively.  These results indicate that there is a finite energy
gap between the ground state and the excited states with respect to
site-located charges and spins.

We can also obtain the bond-bond correlation functions as $\bigl\langle
B^\sigma_{i,i+1} B^{\pm\sigma}_{j,j+1} \bigr\rangle = \bigl(1 \pm
(-1)^{|i-j|}\bigr)/2$, and thus we obtain the {BSDW-$z$} correlation as
$\bigl\langle {\cal O}_{i}^{z}{\cal O}_{j}^{z} \bigr\rangle = 1$.

This result shows the existence of the {BSDW-$z$} long-range order, and
the expectation values of the {BSDW-$z$} order parameter is given by
$\bigl\langle {\cal O}_{i}^{z} \bigr\rangle = \pm 1$.  Since
$\{\,s_{i,i+1,\sigma}^{\mathstrut}, a_{j,j+1,\sigma}^{\dagger}\} = 0$,
the expectation value of ${\cal O}_{i}^{z}$ can be expressed as
$\bigl\langle {\cal O}_{i}^{z} \bigr\rangle = \bigl\langle(-1)^{i}(
M_{i\up}-M_{i\dn})\bigr\rangle$ with
$M_{i\sigma}=s_{i,i+1,\sigma}^{\dagger} s_{i,i+1,\sigma}^{\mathstrut}$.
Now we introduce a bond-spin vector operator given as $\bm{T}_{i} =
\frac{1}{2} \sum_{\sigma\sigma'} s_{i,i+1,\sigma}^{\dagger}
\bm{\tau}_{\sigma\sigma'}^{\mathstrut} s_{i,i+1,\sigma'}^{\mathstrut}$.
Then we have
\begin{equation}
  \bigl\langle {\cal O}_{i}^{z} \bigr\rangle = 
  2\bigl\langle (-1)^{i}\,T_{i}^{z} \bigr\rangle.  
\end{equation}
The {BSDW-$z$} order just corresponds to the N\'eel order of the
bond-located spins.  Since the nearest-neighbor commutation relations of
the bond-spin operators are different from those of the ordinary spin
operators, the bond-spin and the spin operators are not exactly
equivalent.

Next we discuss the relation between the BN\'eel and the BSDW states.
As was discussed above, the BN\'eel state has both charge and spin gaps,
whereas the BSDW state has gapped charge and {\it gapless} spin
excitations~\cite{Japaridze}.  Let us consider the Hamiltonian with an
SU(2) symmetry in spin space given at $X=0$, $\Vbot=\Vpara=V $ in
Eq.~(\ref{eqn:ghub-l}).  In this Hamiltonian,
$\bigl|\Psi_{0}^{\sigma}\bigr\rangle$ gives an exact ground state when
$U=2V=2W=2t$.  There is large degeneracy in the ground state and the
correlation function of the BN\'eel state are not reliable.

\begin{figure}
 \begin{center}
\noindent \epsfxsize=2.9in \leavevmode \epsfbox{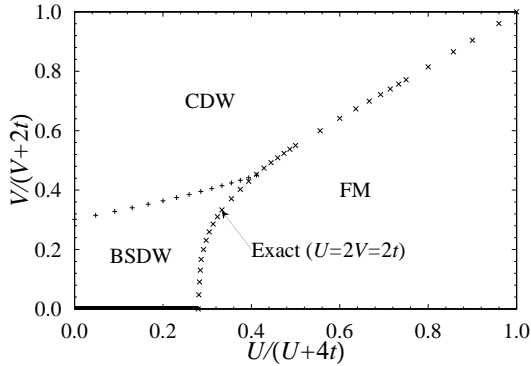}
 \end{center}
\caption{The phase diagram of the model in the SU(2) symmetric case with
 $X=0,W=t$, obtained by the numerical data of the $L=12$ system. The
 wave function of Eq.~(\protect{\ref{eqn:twf}}) gives the exact ground
 state at $U=2V=2t$, which corresponds to the point indicated in
 Fig.~\protect{\ref{fig:pd1}}.}
\label{fig:pd2}
\end{figure}

We show in Fig.~\ref{fig:pd2} the phase diagram at $W/t=1$ in the
$U$--$V$ plain, obtained by the exact diagonalization of the $L=12$
system.  There appear BSDW, charge-density-wave (CDW), and FM phases.
The CDW state has gaps both in charge and spin excitations, and shows a
site-long-range order.  The spin-gap transition occurs between the
CDW-BSDW boundary.  This transition point is obtained by the
singlet-triplet level crossing in the excited states, which is justified
by the conformal field theory~\cite{Nakamura}.  On the other hand, the
CDW-FM and the BSDW-FM transitions are of the first order, and their
boundaries are determined by level crossing in the ground
state~\cite{Nakamura-I-M}.  Finite size effects in this analysis are
small enough to get reliable results.  The exact ground state appears
just on the BSDW-FM boundary within the precision of the exact
diagonalization.

As was shown in Fig.~\ref{fig:spin_exs}, at the exactly solvable point
($U=2t$), all the lowest states with the total spin $S_{\rm tot}$
($S_{\rm tot}\leq L/2$) degenerate.  From the above numerical and exact
results, we conjecture that this total spin-$S_{\rm tot}$ states can be
constructed by means of linear combinations of the
states~(\ref{eqn:d-states}) and the singlet state in the constructed
states corresponds to the BSDW state.

Finally, we comment on the relation between the present result and other
works.  Exact dimerized ground states in electron systems is discussed
recently by Dmitriev {\it et al.}~\cite{Dmitriev-K-O}.  The dimer
consists of up- and down-spin electrons, whereas our ``dimer'' given in
Eq.~(\ref{eqn:twf}) of an electron and a hole for each spin.  Their
results, therefore, do not contain a state with the staggered
magnetization on the bonds.  The ground state we have discussed is
rather similar to the one which appears in a spin-$1/2$ two-leg ladder
model~\cite{Kolezhuk-M} or in an orbitally degenerate spin
chain~\cite{KI} with the Jordan-Wigner transformation.

In summary, we have shown that a BN\'eel state with the BSDW order is
the exact ground state in wide parameter space of the generalized
Hubbard chain.  In this state, both charge and spin excitations have
gaps.  In the case when SU(2) symmetry exists in the spin space, the
ground state degenerates with higher spin states including the fully FM
state.  This BN\'eel phase adjoins the critical BSDW phase on the
multicritical point.  We expect that the relation between the BN\'eel
and the BSDW states is analogous to the one between the Ising limit and
the Heisenberg point in the spin-$1/2$ antiferromagnetic XXZ chain with
respect to the low-energy spin excitations.  Note, in the model
(\ref{eqn:ghub-l}), that the BN\'eel state appears when the spin
anisotropy is XY-like, $\Vbot/W=\Vbot/\Vpara<1$, contrary to the N\'eel
state of the XXZ model.


\end{document}